\documentclass[a4paper]{iopart}
\usepackage{iopams}
\expandafter\let\csname equation*\endcsname=\relax 
\expandafter\let\csname endequation*\endcsname=\relax 
\usepackage[dvipdfmx]{graphicx,hyperref}
\usepackage{amsmath,amssymb,dcolumn,float,enumitem,bm,comment,cases,empheq,mathrsfs,psfrag}
\def\Introduction{Introduction}

\def\FormulationA{Thermodynamic quantities of a $D$-dimensional Bose gas}
\def\FormulationB{Landau's two-fluid model}
\def\SoundvThreeD{Three-dimensional Bose gas}
\def\SoundvTwoD{Two-dimensional Bose gas}
\def\SoundvOneD{One-dimensional Bose gas}
\def\AppA{Sound velocity in the phononic regime}

\newcommand{\beq}{\begin{equation}}
\newcommand{\eeq}{\end{equation}}
\newcommand{\bal}{\begin{aligned}[b]}
\newcommand{\eal}{\end{aligned}}

\begin{document}

\title{Sound modes in collisional superfluid Bose gases}
\author{K. Furutani$^{1}$, A. Tononi$^{1}$, and L. Salasnich$^{1,2,3}$}
\address{$^{1}$ Dipartimento di Fisica e Astronomia ``Galileo Galilei'', 
Universit\`a di Padova, via Marzolo 8, 35131 Padova, Italy. \\
$^{2}$ CNR-INO, via Nello Carrara, 1 - 50019 Sesto Fiorentino, Italy.\\
$^{3}$ Istituto Nazionale di Fisica Nucleare (INFN), Sezione di Padova, 
via Marzolo 8, 35131 Padova, Italy}
\begin{abstract}
We theoretically investigate sound modes in a weakly-interacting collisional Bose gas in $D$ dimensions. Using the Landau's two-fluid hydrodynamics and working within the Bogoliubov theory, we observe the hybridization of the first and second sound modes for $D\ge 2$. 
To model the recent measurements of the sound velocities in 2D, obtained in the weakly-interacting regime and around the Berezinskii-Kosterlitz-Thouless transition temperature, we derive a refined calculation of the superfluid density, finding a fair agreement with the experiment.
In the 1D case, for which experimental results are currently unavailable, we find no hybridization, triggering the necessity of future investigations. 
Our analysis provides a systematic understanding of sound propagation in a collisional weakly-interacting Bose gas in $D$ dimensions. 
\end{abstract}
\maketitle

\section{\Introduction}\label{SecIntro}

The low-temperature physics of quantum liquids, whose study ranges from the seminal experiments of Kapitza with liquid Helium \cite{kapitza} to the developments in the field of ultracold quantum gases \cite{pethick,pitaevskii}, is constructed on the paradigm of superfluidity. 
This quantum mechanical phenomenon, usually defined as the capability of a quantum liquid to ``flow without friction'' through narrow capillaries, has important observable consequences on the dynamical properties of quantum liquids. 
As a corollary, the study of dynamical phenomena, for instance the propagation of sound, provides nontrivial information on the superfluid character of the system, and on its thermodynamical and near-to-equilibrium properties. 
Specifically, depending on the physical regime defined by the parameter $\omega \tau$, where $\omega$ is the sound-wave frequency, and $\tau$ is the mean time between each collision, sound propagation occurs in different qualitative ways. 

When $\omega \tau \gg 1$ the collisions between the atoms are rare and the propagation of collisionless sound originates from the mean-field interaction of the fluid. 
Historically, Andreev and Khalatnikov studied the propagation of sound in this regime \cite{andreev}, explaining previous experiments with liquid $^4$He \cite{whitney}, but more recently, collisionless sound has been the object of renewed experimental and theoretical interest in two-dimensional ultracold atomic gases \cite{ville,ota,cappellaro}. 

In this work, however, we will focus on the propagation of sound in collisional superfluid Bose gases, and we will consider the collisional regime of $\omega \tau < 1$. In this case, the hydrodynamic properties of a $D$-dimensional Bose gas can be described with Landau and Tisza two-fluid model \cite{landau, landaufluid}, in which the quantum fluid is described as a mixture of a normal component and a superfluid component. While the normal part of the fluid is viscous, the superfluid one flows without friction and does not carry entropy. As a consequence in the near-to-equilibrium dynamics, due to the existence of these two macroscopic degrees of freedom, a second sound mode appears alongside the ``usual'' first one. 
While the two-fluid model is a general framework, valid both in bosonic and fermionic systems, and in different spatial dimensions, the microscopic mechanisms underlying the qualitative and quantitative physical description of the first and second sound are system-dependent. 

Due to the large isothermal compressibility of a 3D weakly-interacting Bose gas, and a similar behaviour is expected to occur in a 2D Bose gas \cite{singh}, the first and second sound modes hybridize \cite{lee, griffin, taylor, hu, verney}. 
This phenomenon reveals the inversion of the role of density and entropy oscillations in the propagation of first and second sound: their contribution to the sound modes exchange when the finite hybridization temperature is crossed. 

In 2D Bose gases, the Mermin-Wagner theorem \cite{merminwagner, hohenberg} rules out the occurrence of long-range order at finite temperature, nonetheless the superfluid density can be finite at temperatures below the Berezinskii-Kosterlitz-Thouless (BKT) critical temperature, $T_{\mathrm{BKT}}$ \cite{bere, kosterlitz, nelson,desbusquois}. Therefore, the proliferation of free vortices at $T>T_{\mathrm{BKT}}$ which leads to a jump of the superfluid density \cite{nelson}, results also in the discontinuity of both first sound and second sound velocities. In the vicinity of the BKT transition temperature, an analysis based on universal relations (UR) in two-dimensional Bose gases \cite{prokofev01, prokofev02, yefsah, hung, rancon} is a valid description of this behavior. Indeed, it has succeeded in predicting the sound velocities quantitatively in the temperature regime near $T_{\mathrm{BKT}}$ \cite{ozawa, miki, hadz}. 
Compared to the 3D and 2D cases, there is few investigation of the first and second sound velocities at finite temperature in 1D. To obtain meaningful results in this case, which deserves a detailed analysis, it is important to establish in which temperature regime a hydrodynamic description is reliable. 

In this paper, we systematically investigate the low-temperature behaviour of sound velocities in a $D$-dimensional weakly-interacting Bose gas. Utilizing the two-fluid hydrodynamics and the Bogoliubov theory, we compute the sound velocities in a collisional Bose gas in $D=1,2,3$. We find that the hybridization, which has been predicted theoretically in a 3D Bose gas \cite{lee, griffin, verney}, can occur for $D\ge 2$. In particular, to obtain reliable results near $T_{\mathrm{BKT}}$ in 2D, we calculate the renormalized superfluid density by developing an improved approach based on Popov theory. Our theoretical results are, in this case, in reasonable agreement with the experimental measurements of Ref.~\cite{hadz}. 
In the 1D case, the calculation of the sound velocities do not exhibit any hybridization and our quantitative predictions await experimental confirmations.

\section{\FormulationA}\label{SecFormA}

We start from the Helmholtz free energy of a weakly-interacting $D$-dimensional Bose gas, which, including the quantum correction at zero temperature, reads (we set $\hbar=k_{\mathrm{B}}=1$ throughout this paper)
\beq
\bal
F&=F_{0}+F_{\mathrm{Q}}+F_{T} \\
&=\frac{g}{2}\frac{N^{2}}{L^{D}}+\frac{1}{2}\sum_{\bm{p}}E_{p}+T\sum_{\bm p}\ln{\left[ 1 - e^{-E_{p}/T}\right]} ,
\eal
\label{freeenergy}
\eeq
where $F_{0}$ is the mean-field zero-temperature free energy with $g$ is the Bose-Bose interaction strength, $N$ is the total number of identical bosons confined in a hypercube of side $L$ and hypervolume $L^{D}$. $F_{T}$ is the low-temperature free energy with $T$ is the absolute temperature and
\beq
E_{p}=\sqrt{\frac{p^{2}}{2m}\left(\frac{p^{2}}{2m}+2gn\right)},
\eeq
is the Bogoliubov spectrum where $n=N/L^{D}$ is the $D$-dimensional number density and $m$ is the mass of the atoms. We define a gas parameter in $D$ dimension as
\beq
\eta\equiv \frac{mgn^{1-2/D}}{2\pi},
\eeq
which is indeed identical to $gn/[T_{\mathrm{c}}\, \zeta(D/2)^{2/D}]$ for $D=3$ where $T_{\mathrm{c}}$ is the critical temperature in the noninteracting case. The quantum correction $F_{\mathrm{Q}}$ in the free energy is obviously ultraviolet divergent and requires a regularization procedure. Dimensional regularization \cite{toigo} for each spatial dimension leads to
\beq
F_{\mathrm{Q}}=
\begin{cases}
\displaystyle L^3 \frac{8}{15\pi^2} m^{3/2} 
\left( g n \right)^{5/2} & \text{($D=3$)}, \\ \\
\displaystyle-L^2 \frac{m}{8\pi} \left[ 
\ln{\left({\epsilon_{\Lambda}\over g n} 
\right)} - \frac{2}{\eta} \right] \, \left(g n\right)^2 
& \text{($D=2$)}, \\ \\
\displaystyle-L \frac{2}{3\pi} m^{1/2} 
\left( g n \right)^{3/2} & \text{($D=1$)} ,
\end{cases}
\label{FQ}
\eeq
where $\epsilon_{\Lambda}=4e^{-2\gamma-1/2}/\left(ma_{\mathrm{2D}}^{2}\right) \gg gn$ \cite{mora} is a cutoff energy for $D=2$ and $\gamma=0.577\cdots$ is the Euler-Mascheroni's constant. The 2D $s$-wave scattering length $a_{\mathrm{2D}}$ is related to the 2D coupling constant as \cite{toigo, dalibard}
\beq
g=\frac{2\pi}{m}\frac{1}{\ln{\left(2/\left(e^{\gamma}ka_{\mathrm{2D}}\right)\right)}} ,
\label{ga2d}
\eeq
within the Born approximation, and, substituting $k=\pi/L$, one can obtain
\beq
\frac{\epsilon_{\Lambda}}{gn}=\frac{\pi }{2N}\frac{e^{2/\eta-1/2}}{\eta} .
\label{cutoff}
\eeq
The pressure $P$ is obtained as
\beq
P=-\left(\frac{\partial F}{\partial L^{D}}\right)_{N,T} .
\eeq
Other thermodynamic quantities can be obtained as well from the Helmholtz free energy in Eq.~(\ref{freeenergy}). The entropy per mass unit $s$ and the specific heat at constant volume $c_V$ are given by
\beq
s=\frac{1}{m}\left(\frac{\partial}{\partial T}\frac{F}{N}\right)_{\rho} ,
\quad\quad
c_{V}=T\left(\frac{\partial s}{\partial T}\right)_{\rho} ,
\eeq
where $\rho=mn$ denotes the mass density.

\section{\FormulationB}\label{SecformB}

The calculation of the thermodynamic functions is usually based on microscopic derivations, as the one outlined in the previous section, that depend on the specific system and on its physical regimes. 
Landau's two-fluid model, in which the system is described as a mixture of a viscous normal fluid and a non-viscous superfluid, is however a general theoretical framework to describe the hydrodynamic properties of quantum liquids.
Within this model, the first sound velocity $u_{1}$ and the second sound velocity $u_{2}$ ($\le u_{1}$) are the solutions of the following biquadratic equation \cite{landau}
\beq
u^{4}-\left(v_{\mathrm{A}}^{2}+v_{\mathrm{L}}^{2}\right)u^{2}+v_{\mathrm{T}}^{2}v_{\mathrm{L}}^{2}=0,
\label{Landaueq}
\eeq
where we define the isothermal, adiabatic, and Landau velocity, respectively,  as \cite{landau, khala}
\beq
v_{\mathrm{T}} = \sqrt{\left(\frac{\partial P}{\partial \rho}\right)_{T}} , 
\quad
v_{\mathrm{A}} = \sqrt{\left(\frac{\partial P}{\partial \rho}\right)_{s}} , 
\quad
v_{\mathrm{L}}=\sqrt{\frac{\rho_{\mathrm{s}}Ts^{2}}{\rho_{\mathrm{n}}c_{V}}} ,
\label{vTAL}
\eeq
and the total mass density $\rho=mn=\rho_{\mathrm{n}}+\rho_{\mathrm{s}}$  is the sum of the normal mass density $\rho_{\mathrm{n}}$ and the superfluid mass density $\rho_{\mathrm{s}}$. Note that the Landau velocity $v_{\mathrm{L}}$ defined here corresponds to the velocity of a pure entropy wave.

The thermodynamic quantities in Eq. (\ref{vTAL}) depend on the system considered: here we implement their calculation following the Bogoliubov theory introduced in the last section, which describes a weakly-interacting Bose gas in $D$ dimensions. Moreover, we calculate the normal mass density as \cite{landau, landaufluid}
\beq
\rho_{\mathrm{n}}=-\frac{1}{D}\int\frac{d^{D}\bm{p}}{(2\pi)^{D}} \, p^{2}\frac{dn_{\mathrm{B}}(E_{p})}{dE_{p}} ,
\label{rhon}
\eeq
which, for a noninteracting gas with $g=0$, reduces to the total mass density $\rho_{\mathrm{n}}=\rho$ and the superfluid fraction vanishes. In Eq. (\ref{rhon}), $n_{\mathrm{B}}(E)=\left[e^{E/T}-1\right]^{-1}$ is the Bose distribution function.

\begin{figure}[t]
\centering
\includegraphics[width=80mm]{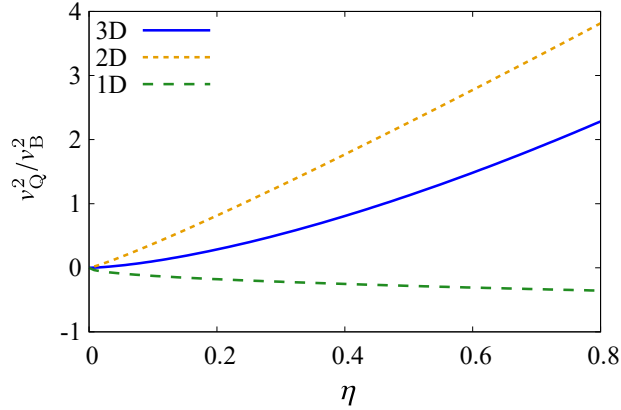}
\caption{The beyond-mean-field correction to the isothermal and adiabatic velocity $v_{\mathrm{Q}}^{2}$ for $D=1,2,3$. The horizontal axis is the gas parameter $\eta= mgn^{1-2/D}/\left(2\pi\right)$. For $D=2$, the number of particles is set to $N=10^{4}$.}
\label{vQ_123D}
\end{figure}

Denoting $\bar{P}$ as the pressure contribution which includes the mean-field plus the thermal one, and $P_{\mathrm{Q}}$ as the quantum correction, one can obtain
\beq
v_{\mathrm{T}}=\sqrt{\left(\frac{\partial\left(\bar{P}+P_{\mathrm{Q}}\right)}{\partial\rho}\right)_{T}}
=\sqrt{\bar{v}_{\mathrm{T}}^{2}+ v_{\mathrm{Q}}^{2}} ,
\eeq
where $\bar{v}_{\mathrm{T}}$ is the isothermal velocity within the mean-field theory and
\beq
v_{\mathrm{Q}}^{2}\equiv \left(\frac{\partial P_{\mathrm{Q}}}{\partial\rho}\right)_{T} ,
\eeq
is the beyond-mean-field correction to the isothermal velocity. Since $F_{\mathrm{Q}}$ is the zero-temperature free energy, it does not affect the Landau velocity $v_{\mathrm{L}}$ and the quantum correction to the adiabatic velocity is identical to that to the isothermal one as
\beq
v_{\mathrm{A}}=\sqrt{\bar{v}_{\mathrm{A}}^{2}+v_{\mathrm{Q}}^{2}} ,
\eeq
where $\bar{v}_{\mathrm{A}}$ is the adiabatic velocity within the mean-field theory. The explicit expressions of the quantum correction $v_{\mathrm{Q}}^{2}$ are given by
\beq
v_{\mathrm{Q}}^{2}=
\begin{cases}
\displaystyle\frac{2\left(2\pi\eta\right)^{3/2}}{\pi^{2}}v_{\mathrm{B}}^{2} & \text{($D=3$)} , \\ \\
\displaystyle-\frac{\eta}{2}\left[\ln{\left(\frac{\epsilon_{\Lambda}}{gn}\right)}-\frac{2}{\eta}-\frac{1}{2}\right]v_{\mathrm{B}}^{2} & \text{($D=2$)} , \\ \\
\displaystyle-\sqrt{\frac{\eta}{2\pi}}v_{\mathrm{B}}^{2}  & \text{($D=1$)} ,
\end{cases}
\eeq
where the Bogoliubov velocity reads $v_{\mathrm{B}}=\sqrt{gn/m}$. Fig. \ref{vQ_123D} represents the quantum correction $v_{\mathrm{Q}}^{2}$ to the gas parameter $\eta$ in each dimension. One can see that $v_{\mathrm{Q}}^{2}$ vanishes as $\eta\to 0$ in any dimension. 
The quantum correction $v_{\mathrm{Q}}^{2}$ is positive in 3D while it is negative in 1D. In 2D, it is positive for $\eta>\pi/\left(2eN\right)$ and in the thermodynamic limit $N\to\infty$, one can assume $v_{\mathrm{Q}}^{2}>0$.

Our theoretical framework is reliable in physical regimes where the hydrodynamic description of the system is valid. 
In particular, it is necessary that $\omega\tau \ll 1$ with $\tau$ is the collisional time and $\omega\simeq v_{\mathrm{B}}k$ is the frequency of the excited phononic mode. The collisional time is given by
\beq
\tau\sim\frac{l_{\mathrm{mfp}}}{v_{\mathrm{th}}}\sim\frac{1}{n\sigma v_{\mathrm{th}}} ,
\eeq
where $l_{\mathrm{mfp}}\sim1/(n\sigma)$ is the mean-free-path and $v_{\mathrm{th}}=\sqrt{2T/m}$ is the thermal velocity. For $D=3$, the cross-section is given by $\sigma=4\pi a^{2}=m^{2}g^{2}/(4\pi)$, which leads to
\beq
\omega\tau\sim N^{-\frac{1}{3}}\frac{\eta^{-2}}{\sqrt{2t}} .
\label{omegatau3D}
\eeq
Equation (\ref{omegatau3D}) indicates that our hydrodynamic description is valid at high temperature, for a large gas parameter, or for a large number of particles. Taking into account the Bogoliubov theory under the low-temperature approximation we employed, our theory would be valid under low temperature, small gas parameter, and a large number of particles.
The cross-section for $D=2$ is given by $\sigma\sim\left(2\pi\eta\right)^{2}/(mv_{\mathrm{th}})$ and the adimensional collisional time is independent of the temperature as
\beq
\omega\tau\sim\frac{1}{2\sqrt{2\pi N}}\eta^{-\frac{3}{2}} .
\label{omegatau2D}
\eeq
Equation (\ref{omegatau2D}) indicates that the hydrodynamic description for $D=2$ is valid for a large gas parameter or a large number of particles. As in 3D case, working with the Bogoliubov theory under the low-temperature approximation, our 2D theory is valid under the conditions of low temperature, small gas parameter, and a large number of particles. In the experimental observation reported in Ref. \cite{hadz}, the gas parameter and the number of particles are $\eta\simeq 0.10$ and $N\simeq2178$ respectively, and one obtains $\omega\tau\simeq0.13$, in which our hydrodynamic description is reliable.

\begin{figure}[t]
\centering
\leavevmode
\begin{minipage}{.49\columnwidth}
\includegraphics[clip=true,height=0.7\columnwidth,width=1\columnwidth]{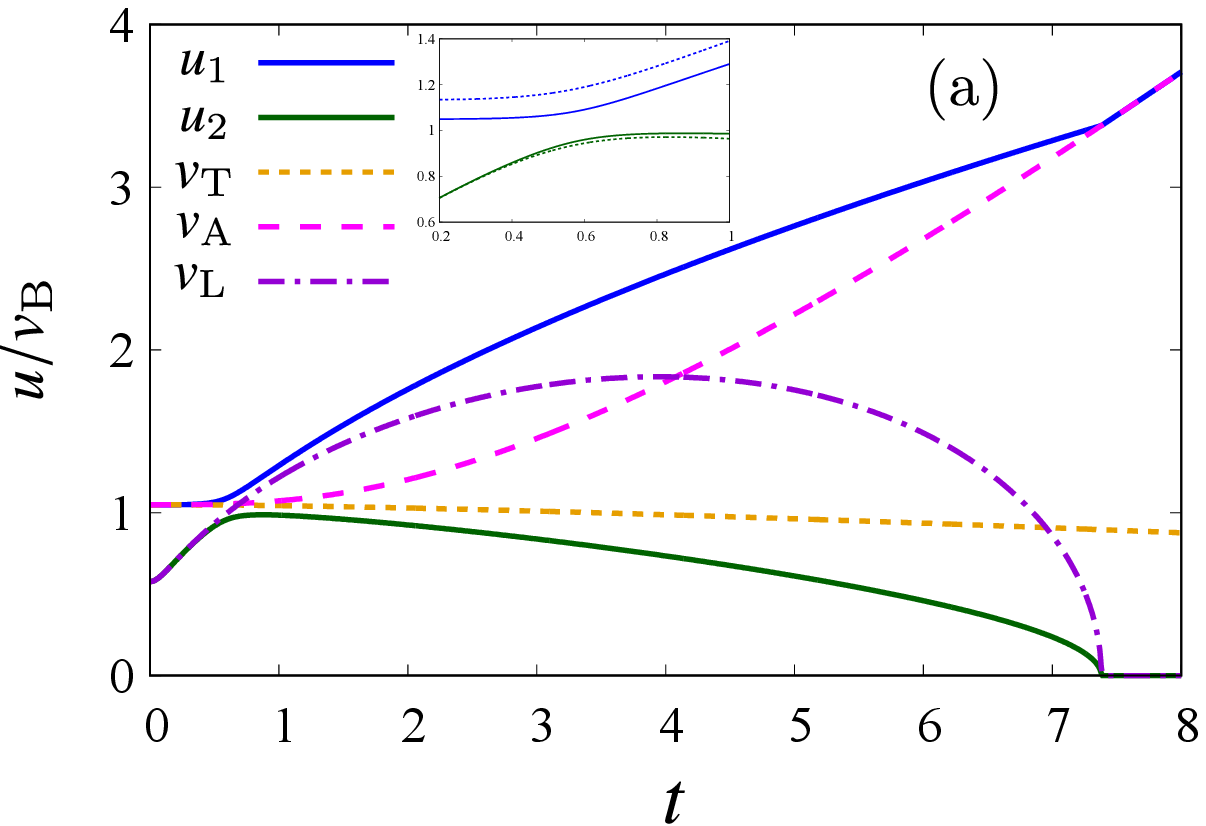}
\end{minipage}
\begin{minipage}{.49\columnwidth}
\includegraphics[clip=true,height=0.7\columnwidth,width=1\columnwidth]{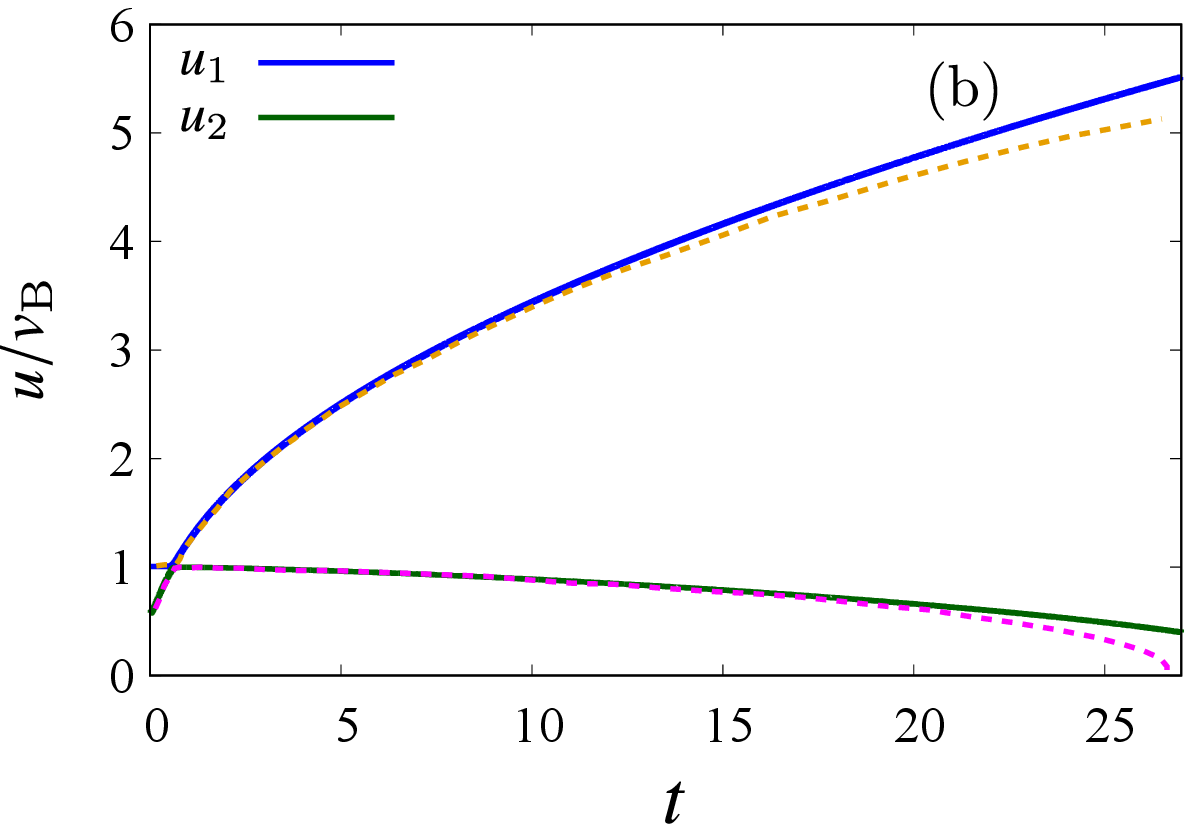}
\end{minipage}
\caption{Results of sound velocities in a weakly-interacting Bose gas for $D=3$ and $\eta=0.1$ (panel (a)) and the comparison with the first and second sound velocity in Ref. \cite{verney} for $\eta=0.02$ (panel (b)). The horizontal axis is the reduced temperature $t=T/(gn)$. Inset of panel (a): The hybridization of the first sound and second sound modes. The dotted lines represent the results for $\eta=0.2$. The dotted lines in panel (b) represent the results of Ref. \cite{verney}.}
\label{u1u2_eta0.1_d=3}
\end{figure}

\subsection{\SoundvThreeD}\label{Soundv_3D}

Let us discuss the propagation of the first sound and second sound in $D=3$. The velocities of these modes are shown in Fig.~\ref{u1u2_eta0.1_d=3}, where the temperature is rescaled as $t=T/(gn)$. Note that, in the left panel of Fig.~\ref{u1u2_eta0.1_d=3}, we set the gas parameter to $\eta\equiv mgn^{1/3}/(2\pi)=0.1$. At $T=0$, as discussed in \ref{Phononic}, we reproduce the well-known result:
\beq
u_{1}=\bar{v}_{\mathrm{T}}=\bar{v}_{\mathrm{A}}=v_{\mathrm{B}} ,
\quad\quad
u_{2}=v_{\mathrm{L}}=\frac{v_{\mathrm{B}}}{\sqrt{3}} ,
\eeq
which is given by the mean-field theory.
Around $t=0.6$, it exhibits a hybridization of the two sound modes with a small gap, which has been pointed out by Refs. \cite{lee, griffin, taylor, hu, verney} for a weakly-interacting 3D Bose gas. 
This phenomenon can be interpreted by using the thermal expansion coefficient $\alpha\equiv -\rho^{-1}\left(\partial\rho/\partial T\right)_{P}=\left(v_{\mathrm{A}}^{2}/v_{\mathrm{T}}^{2}-1\right)/T$ in the following way. In the incompressible regime $\alpha T\ll 1$, the biquadratic Landau equation of Eq. (\ref{Landaueq}) gives $u_{1}=v_{\mathrm{A}}$ and $u_{2}=v_{\mathrm{L}}$, which indicates that the first sound and second sound mode correspond to the density mode and the entropy mode respectively. The hybridization temperature $t_{\mathrm{hyb}}$ characterizes this incompressible regime as $t \lesssim t_{\mathrm{hyb}}$. Experimentally, above the hybridization temperature, the second sound can be probed by a density perturbation while only the first sound can be probed below the hybridization temperature since the second sound corresponds to the entropy mode uncoupled from the density oscillation. At higher temperature than the critical temperature at which the the Landau velocity vanishes, one can check that the first sound velocity coincides with the adiabatic one $u_{1}=v_{\mathrm{A}}$. The inset of the left panel shows the first sound and second sound velocities for $\eta=0.1$ and $\eta=0.2$. It exhibits that a larger gas parameter opens the gap larger as $\eta^{3/4}$ \cite{verney}. In 3D, the hybridization occurs for any gas parameters.

The dotted, dashed, and dotted-dashed line in the left panel of Fig. \ref{u1u2_eta0.1_d=3} indicate the isothermal, adiabatic, and Landau velocity for $D=3$ respectively. The normal density fraction $\rho_{\mathrm{n}}$ within the Landau's prescription does not include effects of interactions among elementary excitations and is a low-temperature approximation. In addition, the Bogoliubov theory is not applicable at high temperature regime comparable with $T_{\mathrm{c}}$, so that the critical temperature at which $\rho_{\mathrm{s}}$ vanishes cannot exactly coincide with the superfluid phase transition temperature $T_{\mathrm{c}}$ \cite{verney}. 

We can also qualitatively reproduce the results of Ref. \cite{verney} as shown in the right panel of Fig.~\ref{u1u2_eta0.1_d=3} while our framework ignored the Lee-Huang-Yang correction \cite{LHY}, which is included in Ref. \cite{verney}. Since Ref. \cite{verney} employed perturbation theory based on Beliaev diagrammatic technique at higher temperature region for better prediction, we find deviations in this region.

\begin{figure}[t]
\centering
\includegraphics[width=80mm]{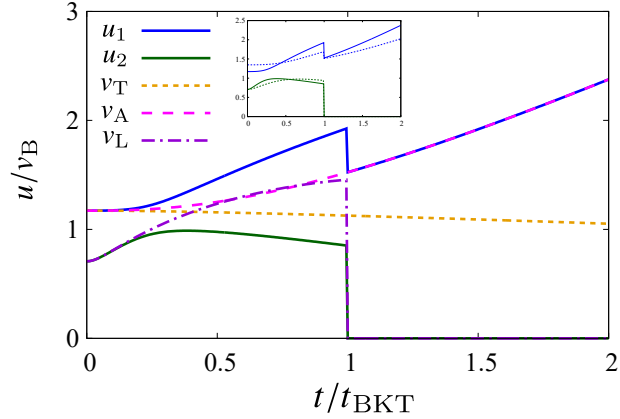}
\caption{Results of sound velocities for $D=2$ and $\eta=0.1$. The number of particle is set to be $N=10^{4}$. The horizontal axis is the reduced temperature scaled by the BKT transition temperature $t_{\mathrm{BKT}}=T_{\mathrm{BKT}}/(gn)$, which is determined by the KT-Nelson's formula in Eq.~\eqref{KTNelson} for the superfluid density in the two-fluid model while $\rho_{\mathrm{n}}$ is computed by Eq. (\ref{rhon}). Inset: The first sound and second sound velocity. The solid lines represent the results for $\eta=0.1$ and the dotted ones represent those for $\eta=0.2$}
\label{u1u2_etat0.1_d2}
\end{figure}

\subsection{\SoundvTwoD}\label{Soundv_2D}

The superfluid properties of a 2D Bose gas are crucially different from those of the 3D case due to the phenomenology of the BKT transition \cite{bere, kosterlitz, nelson}. The theoretical framework developed in Sec.~\ref{SecFormA}, where the topological excitations of the bosonic fluid are not taken into account, cannot describe the BKT transition. These excitations are responsible for the universal jump of the superfluid density at BKT transition temperature, $T_{\mathrm{BKT}}$. To include it in our theory, we employ the KT-Nelson's formula \cite{nelson}
\beq
\frac{\pi}{2m^{2}}\rho_{\mathrm{s}}=T_{\mathrm{BKT}} ,
\label{KTNelson}
\eeq
which determines the BKT transition temperature $T_{\mathrm{BKT}}$. The superfluid density $\rho_{\mathrm{s}}$ in Eq. (\ref{KTNelson}) is calculated from the Landau formula given in Eq. (\ref{rhon}). A good approximation in an infinite-size weakly-interacting system is to set to zero the superfluid density fraction for $T\ge T_{\mathrm{BKT}}$.

\begin{figure}[t]
\centering
\includegraphics[width=80mm]{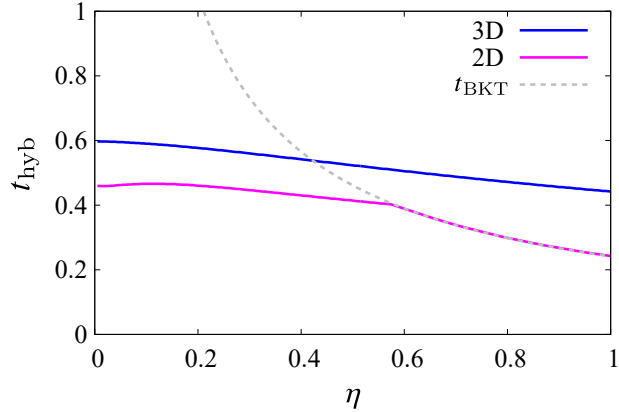}
\caption{Hybridization temperature for $D=3$ and $D=2$ as a function of the gas parameter $\eta= mgn^{1-2/D}/\left(2\pi\right)$. In the latter case, the particle number is set to $N=10^{4}$. In 2D, moreover, the hybridization temperature coincides with the BKT transition temperature for $\eta\gtrsim 0.6$.}
\label{thyb2D}
\end{figure}

We show the sound velocities in a 2D Bose gas in Fig. \ref{u1u2_etat0.1_d2}. Due to the jump of the superfluid density at $t=t_{\mathrm{BKT}}$, the first sound and second sound velocity exhibit discontinuities. One can see that the hybridization of $u_{1}$ and $u_{2}$ occurs around $t_{\mathrm{hyb}}\simeq 0.4$ for $\eta=0.1$. 
Fig. \ref{thyb2D} displays the dependence on the gas parameter $\eta$ of the hybridization temperature $t_{\mathrm{hyb}}$, which is determined by the temperature at which the difference between the first and second sound velocity starts to increase.
Note that in 2D, for $\eta\gtrsim0.6$, $t_{\mathrm{hyb}}$ coincides with the BKT transition temperature. In the the region of $\eta\gtrsim 0.6$, at which $t_{\mathrm{hyb}}=t_{\mathrm{BKT}}$ in 2D, we infer from Fig. \ref{thyb2D} that the first and second sound modes are decoupled, respectively, to density and entropy modes, because the first sound corresponds to the density mode $u_{1}=v_{\mathrm{A}}$ and the second sound vanishes $u_{2}=0$ in the absence of the superfluid density above $t_{\mathrm{BKT}}$.

Our theoretical approach, based on the Bogoliubov theory, is reliable to describe the propagation of sound in low-temperature Bose gases, and its predictions are as better as the gas parameter $mg$ is smaller than 1. 
The recent experiments of Ref.~\cite{hadz} with 2D weakly-interacting bosonic superfluids adopt the value of $mg=0.64$, and, therefore, can be described with our Bogoliubov theory. 
However, since these experiments focus on the high-temperature regime near $T_{\text{BKT}}$, it is useful to extend our previous results to improve the agreement in this specific temperature regime. 

In particular, the sound velocities are strongly dependent on the superfluid density and the one derived from the Landau formula of Eq.~\eqref{rhon} has, strictly speaking, a simplified behaviour near $T_{\text{BKT}}$.

To improve our theory in the high-temperature regime of the experiments we evaluate the renormalized superfluid density, $\rho_{s}^{\mathrm{(R)}}$, by solving the Nelson-Kosterlitz renormalization group equations \cite{nelson}. These differential equations describe the renormalization of the superfluid density due to the presence of vortex-antivortex excitations, which are not taken into account by the Landau formula of Eq.~\eqref{rhon}. They read \cite{nelson}
\beq
\begin{gathered}
\bal
&\partial_l\, K^{-1}(l) = 4 \pi^3 y^2(l) ,\\
&\partial_l\, y(l) = [2-\pi K(l)]\, y(l) ,
\label{rg}
\eal
\end{gathered}
\eeq
where $K(l)=\rho_{\mathrm{s}}(l)/(mT)$, with $\rho_{\mathrm{s}}(l)$ the superfluid density at the adimensional scale $l$, and $y(l)=\exp\left[-\mu_{\mathrm{c}}(l)/T\right]$ is the fugacity, where $\mu_{\mathrm{c}}(l)$ is the vortex chemical potential at scale $l$. 

To describe consistently the finite-size experiments, we solve numerically these equations up to a finite scale, $l_{\mathrm{max}} = \ln(A^{1/2}/\xi)$, where $A$ is the area of the system and $\xi=(g \rho)^{-1/2}$ is the healing length, corresponding approximately to the vortex core size. 
In the solution of Eqs.~\eqref{rg} the choice of the initial conditions is quite delicate: we choose the chemical potential of the bare vortices as $\mu_{\mathrm{c}}(0) = \pi^2 \rho_{\mathrm{s}}(0) /(2m^{2})$ \cite{bighin}, and for the initial value of $K(0)$ we use $K(0)=\rho_{\mathrm{s}}(0)/(mT)$, with $\rho_{\mathrm{s}}(0) = \rho- \rho_{\mathrm{n}}(0)$. It is important to point out that the bare Landau density which we introduce here, $\rho_{\mathrm{n}}(0)$, is formally the same as Eq.~\eqref{rhon}, but is calculated with the Popov spectrum: 
\beq
E_{\mathrm{Pop},p}=\sqrt{\frac{p^{2}}{2m}\left(\frac{p^{2}}{2m}+2\mu\right)},
\eeq
where $\mu$ is the chemical potential of the system. 
We derive this chemical potential as a function of $N$ and of $T$ by inverting numerically the grand canonical equation of state, which reads (see Ref.~\cite{mora})
\begin{equation}
N = \frac{m\mu L^D}{4\pi} \ln\bigg( \frac{4}{m\mu a_{\mathrm{2D}}^2e^{2\gamma+1}} \bigg) + \sum_{\bm{p}} \frac{p^2}{2m} \frac{n_{\mathrm{B}}(E_{\mathrm{Pop},p})}{E_{\mathrm{Pop},p}}.
\label{N}
\end{equation}
In particular, we evaluate $a_{\mathrm{2D}}$ as \cite{dalibard}
\begin{equation}
a_{\mathrm{2D}} = 2.092 \, a_{z} \ln \bigg(-\sqrt{\frac{\pi}{2}} \frac{a_{z}}{a_{\mathrm{3D}}}  \bigg),
\end{equation}
where $a_{z}$ is the characteristic length of the transverse harmonic confinement and $a_{\mathrm{3D}}$ is the three-dimensional $s$-wave scattering length, which is directly controlled in the experiment \cite{hadz}. 
The procedure described above allows us to have reliable results near $T_{\text{BKT}}$ for $\rho_{\mathrm{s}}^{\mathrm{(R)}}\equiv\rho_{\mathrm{s}}(l_{\mathrm{max}})$. 
Given the renormalized density $\rho_{\mathrm{s}}^{\mathrm{(R)}}$ for every temperature $T$, we use it as an input to calculate the sound velocities. 

\begin{figure}[t]
\centering
\leavevmode
\begin{minipage}{.49\columnwidth}
\includegraphics[clip=true,height=0.7\columnwidth,width=1\columnwidth]{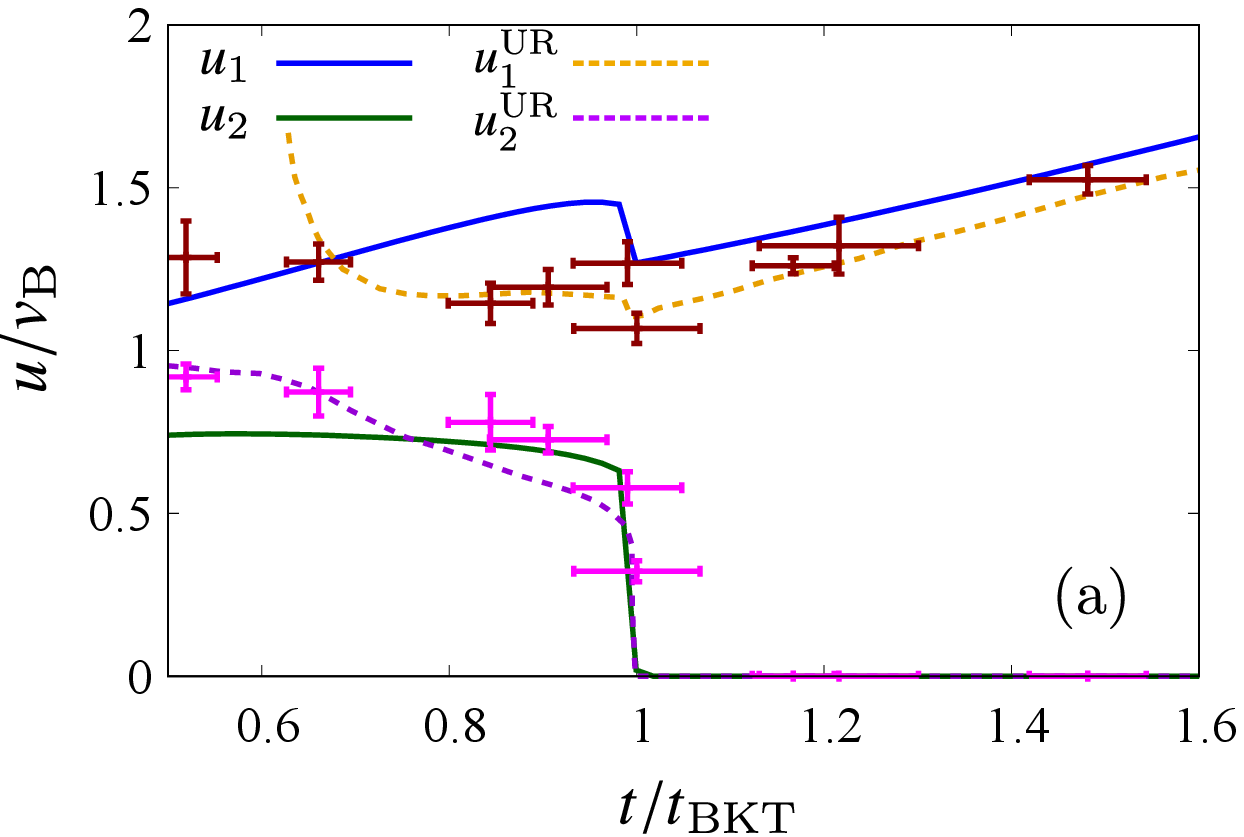}
\end{minipage}
\begin{minipage}{.49\columnwidth}
\includegraphics[clip=true,height=0.7\columnwidth,width=1\columnwidth]{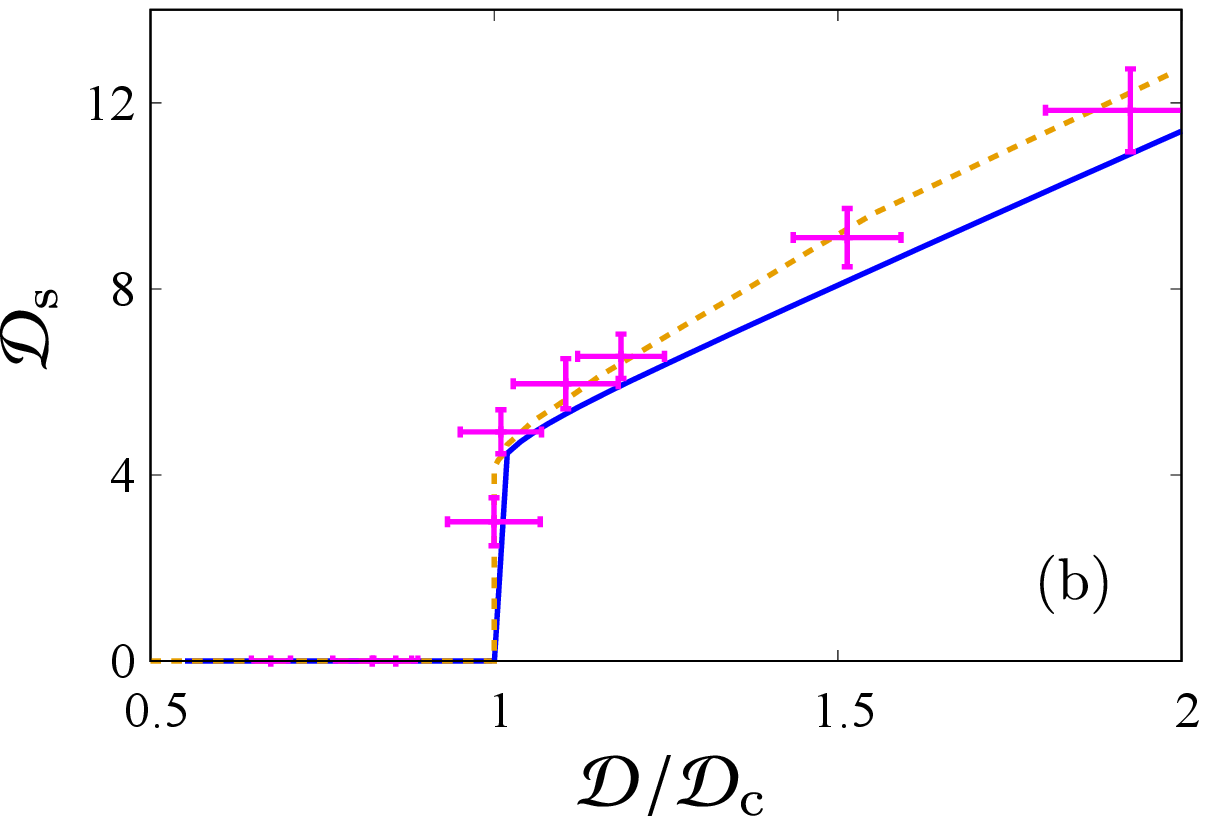}
\end{minipage}
\caption{First sound and second sound velocity (panel (a)) and rescaled superfluid density $\mathcal{D}_{\mathrm{s}}=n_{\mathrm{s}}\lambda_{T}^{2}$ in 2D (panel (b)) for $mg=0.64$, plotted in comparison with the experimental data of Ref.~\cite{hadz} where $\mathcal{D}/\mathcal{D}_{\mathrm{c}}=t_{\mathrm{BKT}}/t$. The particle number is set to be $N=2178$ \cite{hadz}. The blue and green solid lines represent our results using the renormalized superfluid density \cite{nelson} calculated with the chemical potential obtained from Eq.~\eqref{N}. The orange and violet dashed lines represent the results of UR analysis \cite{prokofev01, prokofev02, yefsah, hung, rancon,ozawa,miki,hadz}.}
\label{2dQ_u1u2}
\end{figure}

Our results are outlined in Fig.~\ref{2dQ_u1u2}, which shows, in comparison with the experimental data \cite{hadz}, $u_{1}$ and $u_{2}$ in the left panel and the superfluid density $\mathcal{D}_{\mathrm{s}}=n_{\mathrm{s}}\lambda_{T}^{2}$, with $\lambda_{T}\equiv\sqrt{2\pi/\left(mT\right)}$ the thermal wavelength, in the right panel. As in the experiment, here we use $mg = 0.64$, the number density of $n = 3 \, \mu \text{m}^{-2}$ and the system area of $A = 33 \times 22 \, \mu\text{m}^2$ \cite{hadz}. We also emphasize that, within our finite-size renormalization group calculation, we find a critical BKT temperature of $37 \, \text{nK}$, which is practically coincident with the result $T_{\mathrm{BKT}}=2\pi n/\left[m\ln{\left[380/\left(mg\right)\right]}\right]$ of Ref.~\cite{prokofev01}, and compatible with the critical temperature of the experiments of $42 \, \text{nK}$ \cite{hadz}. Fig. \ref{2dQ_u1u2}(a) indicates that the results using the renormalized superfluid density fraction with the exact chemical potential, represented by the blue and green solid line, are in reasonable agreement with the experimental values. Note that our first sound velocity also describes the behaviour towards low temperature in a satisfactory way. The slight deviation of our second sound velocity from the experimental one at low temperature is ascribed to the inconsistency between the thermodynamic quantities that appear in Eq. (\ref{vTAL}), calculated under the low-temperature approximation $\mu=gn$, and the renormalized superfluid density $\rho_{\mathrm{s}}^{\mathrm{(R)}}$ calculated with the improved $\mu$. While this approximation is not particularly problematic near $T_{\mathrm{BKT}}$, it does not allow us to extend the present theory at low temperatures, where the sound velocities are more sensitive to the normal fraction. Fig.~\ref{2dQ_u1u2}(b) displays that the renormalized $\mathcal{D}_{\mathrm{s}}$ obtained with the beyond-mean-field chemical potential agrees well with the experimental values. From this last figure, thus, we can expect that the corrections due to the interaction between Bogoliubov quasiparticles, which will be more relevant in the high temperature regime and outside the very weakly-interacting regime of $mg \ll 1$, are, at least for the superfluid density, not particularly relevant. This suggests that future works in 2D with the full evaluation of the improved thermodynamics could be a solid benchmark for the sound velocities both in the low and high temperature regimes.

\subsection{\SoundvOneD}\label{Soundv_1D}

On the basis of the Mermin-Wagner theorem \cite{merminwagner}, the critical 
temperature $T_{\mathrm{BEC}}$ below which there is Bose-Einstein condensation, 
or equivalently below which there is off-diagonal long-range order (ODLRO), 
is positive in 3D, it is zero in 2D, and it is absent in 1D. Instead, the 
critical temperature $T_{\mathrm{c}}$ below which there is superfluidity, or 
equivalently below which there is algebraic long-range order (ALRO), 
is equal to $T_{\mathrm{BEC}}$ in 3D, it is equal to $T_{\mathrm{BKT}}$ in 2D, and it is zero 
in 1D. Thus, in the thermodynamic limit and with $T>0$, for a 1D 
weakly-interacting Bose gas there is neither ODLRO nor ALRO. However, 
a finite 1D system of spatial size $L$ is effectively superfluid \cite{super-book} 
if $T\ll E_{\phi}/ \ln{(L/\xi)}$, or equivalently $t \ll t_{\phi}\equiv1/\left[\sqrt{\pi\eta}\ln{\left(2N\sqrt{\pi\eta}\right)}\right]$, where $E_{\phi}\simeq n/(m\xi)$ is the energy to create a phase slip (black soliton) and $\xi$ is the corresponding healing length. Note that, for $\eta\ll 1$, the adimensional temperature $t_{\phi}$ can be quite large.

\begin{figure}[t]
\centering
\includegraphics[width=80mm]{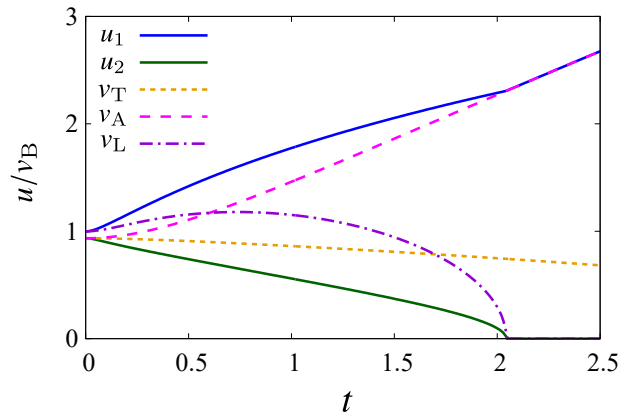}
\caption{Results of sound velocities for $D=1$ and $\eta=0.1$. 
The horizontal axis is the reduced temperature $t=T/(gn)$. }
\label{u1u2_eta0.1_d=1}
\end{figure}

In Fig. \ref{u1u2_eta0.1_d=1}, we show 
the results of $u_{1}$ and $u_{2}$ in an 1D Bose gas for $\eta=0.1$. Since the Bogoliubov theory in 1D well describes the thermodynamics in the weakly-interacting regime up to $\eta\sim 1$ at low temperature \cite{rosi,cappe}, our 1D results would be reliable in this regime. The figure exhibits 
no hybridization of $u_{1}$ and $u_{2}$ because of $u_{1}=u_{2}=v_{\mathrm{B}}$ at zero-temperature 
within the mean-field and the gap opening at zero-temperature between $u_{1}$ and $u_{2}$ by the quantum correction. In the incompressible regime within the mean-field theory, one can obtain $u_{1}=u_{2}=v_{\mathrm{B}}$ in 1D, which indicates that the decoupled density mode and entropy mode are degenerated. Hence, $t=0$ corresponds to the hybridization temperature at which the first and second sound mode are closest to each other in 1D. The beyond-mean-field correction decreases $v_{\mathrm{A}}$ and results in $u_{1}=v_{\mathrm{L}}$ and $u_{2}=v_{\mathrm{A}}$, namely the first and second sound correspond to entropy and density mode respectively due to the negative quantum correction $v_{\mathrm{Q}}^{2}<0$ unlike 3D or 2D case.

\section{Conclusion}

We discuss sound modes in a collisional Bose gas in $D$ dimensions by means of the Landau's two-fluid hydrodynamics and of the Bogoliubov theory. We observe the  hybridization between the first and second sound for $D\ge 2$ and we find that, for a 3D Bose gas in particular, it occurs for any gas parameter.
For 2D collisional superfluids, comparing our theory with the experimental observations of Ref.~\cite{hadz}, we find that, after an improved calculation of the renormalized superfluid density based on the beyond-mean-field equation of state, our results are in fair qualitative agreement with the measured values.
Notably, our results for the superfluid density reproduce quite well the experimental data, suggesting that an improved theory, totally based on the equation of state of Eq.~\eqref{N}, is a promising approach to derive the whole finite-temperature thermodynamics. 
This general calculation, which is expected to reproduce the first and second sound of 2D superfluids quantitatively better, is left for future works.
Finally, we computed the sound velocities also in 1D, finding no hybridization. Since there are no available experimental data yet in this configuration, our 1D analysis could provide a benchmark for future investigations.

\ack
We thank J. Schmitt, M. Ota, and V. Singh for providing the experimental and theoretical data. We also acknowledge valuable comments from M. Ota. KF acknowledges Fondazione Cassa di Risparmio di Padova e Rovigo for a PhD scholarship.

\appendix
\section{\AppA}\label{Phononic}

At very low temperature, according to Ref. \cite{landau}, and under the condition $g\neq 0$, all the thermodynamic functions are fixed by the thermal excitations of the phonons and consequently the Bogoliubov spectrum can be approximated as
\beq 
E_{p} = v_{\mathrm{B}}p .
\label{phonon}
\eeq 
In this phononic regime, within the framework of mean-field theory, one can perform the momentum integral in Eq. (\ref{freeenergy}) analytically as
\beq
\frac{F}{N} =\frac{g\rho}{2m}- \frac{\Omega_{D} \Gamma(D)\zeta(D+1)}{(2\pi)^{D}}\frac{m^{D+1}}{g^{D/2} \rho^{D/2+1}}T^{D+1} , 
\label{Fphononic}
\eeq
\beq
P=\frac{g\rho^{2}}{2m^{2}}+\left(\frac{D}{2}+1\right)\frac{\Omega_D \Gamma(D)\zeta(D+1)}{(2\pi)^{D}} 
\frac{m^{D}}{g^{D/2} \rho^{D/2}}T^{D+1},
\label{Pphononic}
\eeq
where $\Omega_{D}=D\pi^{D/2}/\Gamma(D/2+1)$ is the volume of the $D$-dimensional unit sphere with $\Gamma(x)$ the Euler gamma function and $\zeta(x)$ the Riemann zeta function. The entropy per mass unit, specific heat, and normal density fraction are obtained as
\beq
s=(D+1) \frac{\Omega_{D}\Gamma(D)\zeta(D+1)}{(2\pi)^{D}}\frac{m^{D}}{g^{D/2} \rho^{D/2+1}}T^D ,
\eeq
\beq
c_{v}=(D+1)D\frac{\Omega_{D}\Gamma(D)\zeta(D+1)}{(2\pi)^D}\frac{m^{D}}{g^{D/2} \rho^{D/2+1}}T^{D} ,
\eeq
\beq
\rho_{\mathrm{n}}=(D+1)\frac{\Omega_{D}\Gamma(D)\zeta(D+1)}{(2\pi)^D}\frac{m^{D+2}}{g^{D/2+1} \rho^{D/2+1}}T^{D+1} .
\eeq
Using Eqs. (\ref{Landaueq}), (\ref{vTAL}), at zero-temperature within the mean-field theory, one obtains
\beq
u_{1}=\bar{v}_{\mathrm{T}}=\bar{v}_{\mathrm{A}}=v_{\mathrm{B}}, \quad u_{2}=v_{\mathrm{L}}=\frac{1}{\sqrt{D}}v_{\mathrm{B}} .
\label{u1u2phononic}
\eeq


\begin{thebibliography}{99}


\bibitem{kapitza} Kapitza P 1938 {\it Nature} {\bf141} 74
\bibitem{pethick} Pethick C J and Smith H 2002 {\it Bose-Einstein Condensation in Dilute Gases} 
(Cambridge: Cambridge University Press)
\bibitem{pitaevskii} Pitaevskii L and Stringari S 2016 {\it Bose-Einstein Condensation and Superfluidity} (Oxford: Oxford University Press)
\bibitem{andreev}  Andreev A and Khalatnikov I M 1963 {\it Sov. Phys. JETP } {\bf 17} 1384 
\bibitem{whitney}   Whitney W M and Chase C E 1962 {\it Phys. Rev. Lett.} {\bf 9} 243
\bibitem{ville} Ville J L, Saint-Jalm R, Le Cerf \'E, Aidelsburger M, Nascimb\'ene S, Dalibard J, and Beugnon J 2018
{\it Phys. Rev. Lett.} {\bf 121} 145301 
\bibitem{ota} Ota M, Larcher F, Dalfovo F, Pitaevskii L, Proukakis N P, and Stringari S 2018 {\it Phys. Rev. Lett.} {\bf 121} 145302 
\bibitem{cappellaro} Cappellaro A,  Toigo F, and Salasnich L 2018 {\it Phys. Rev. A} {\bf 98} 043605 
\bibitem{landau} Landau L D 1941 {\it J. Phys. (USSR)} {\bf 5} 71 
\bibitem{landaufluid} Landau L D and Lifshitz E M 1987 {\it Fluid Mechanics} (Oxford: Pergamon)
\bibitem{singh} Singh V P and Mathey L arXiv:2010.00013
\bibitem{lee} Lee T D and Yang C N 1959 {\it Phys. Rev.} {\bf 113} 1406
\bibitem{griffin} Griffin A and Zaremba E 1997 {\it Phys. Rev. A} {\bf 56} 4839
\bibitem{taylor} Taylor E, Hu H, Liu X -J, Pitaevskii L P, Griffin A, and Stringari S 2009 {\it Phys. Rev. A} {\bf 80} 053601
\bibitem{hu} Hu H, Taylor E, Liu X -J, Stringari S, and Griffin A 2010 {\it New. J. Phys.} {\bf 12} 043040
\bibitem{verney} Verney L, Pitaevskii L, and Stringari S 2015 {\it EPL} {\bf 111} 40005 
\bibitem{merminwagner} Mermin N D and Wagner H 1966 {\it Phys. Rev. Lett.} {\bf 17} 1133
\bibitem{hohenberg} Hohenberg P C 1967 {\it Phys. Rev.} {\bf 158} 383
\bibitem{bere} Berezinskii V L 1972 {\it Sov. Phys. JETP} {\bf 34} 610
\bibitem{kosterlitz} Kosterlitz J M and Thouless D J 1972 {\it J. Phys. C} {\bf 5} L124 ; 1973 {\it J. Phys. C} {\bf 6} 1181
\bibitem{nelson} Nelson D R and Kosterlitz J M 1977 {\it Phys. Rev. Lett.} {\bf 39} 1201
\bibitem{desbusquois}     Desbuquois R, Chomaz L, Yefsah T, L\'eonard J, Beugnon J, Weitenberg C, and Dalibard J 2012 {\it Nature Phys.} {\bf 8} 645
\bibitem{prokofev01} Prokof'ev N, Ruebenacker O, and Svistunov B 2001 {\it Phys. Rev. Lett.} {\bf 87} 270402
\bibitem{prokofev02} Prokof'ev N and Svistunov B 2002 {\it Phys. Rev. A} {\bf 66} 043608
\bibitem{yefsah} Yefsah T, Desbuquois R, Chomaz L, G$\rm\ddot{u}$nter K J, and Dalibard J 2011 {\it Phys. Rev. Lett.} {\bf 107} 130401
\bibitem{hung} Hung C -L, Zhang X, Gemelke N, and Chin C 2011 {\it Nature} {\bf 470} 236
\bibitem{rancon} Ran{\c{c}}on A and Dupuis N 2012 {\it Phys. Rev. A} {\bf 85} 063607
\bibitem{ozawa} Ozawa T and Stringari S 2014 {\it Phys. Rev. Lett.} {\bf 112} 025302
\bibitem{miki} Ota M and Stringari S 2018 {\it Phys. Rev. A} {\bf 97} 033604
\bibitem{hadz} Christodoulou P, Ga\l ka M, Dogra N, Lopes R, Schmitt J, and Hadzibabic Z arXiv:2008.06044
\bibitem{toigo} Salasnich L and Toigo F 2016 {\it Phys. Rep.} {\bf 640} 1
\bibitem{mora} Mora C and Castin Y 2009 {\it Phys. Rev. Lett.} {\bf 102} 180404
\bibitem{dalibard} Dalibard J 2016 {\it Fluides quantiques de basse dimension et transition de Kosterlitz-Thouless} (Coll\`ege de France Lecture Notes)
\bibitem{khala} Khalatnikov I M 1965 {\it An Introduction to the Theory of Superfluidity} (New York: Benjamin)
\bibitem{LHY} Lee T D, Huang K and Yang C N 1957 {\it Phys. Rev.} {\bf 106} 1135
\bibitem{bighin} Bighin G and Salasnich L 2017 {\it Sci. Rep.} {\bf 7} 45702
\bibitem{super-book} Svistunov B, Babaev E, and Prokof'ev N 2015 {\it Superfluid States of Matter} (Boca Raton: CRC Press)
\bibitem{rosi} De Rosi G, Astrakharchik G E, and Stringari S 2017 {\it Phys. Rev. A} {\bf 96} 013613
\bibitem{cappe} Cappellaro A and Salasnich L 2017 {\it Phys. Rev. A} {\bf 96} 063610
\end{thebibliography}
\end{document}